\documentstyle[12pt,psfig,oldlfont]{article}

\font\tenrm=cmr10

\begin{document}

\renewenvironment{thebibliography}[1]
  { \begin{list}{\arabic{enumi}.}
    {\usecounter{enumi} \setlength{\parsep}{0pt}
     \setlength{\itemsep}{3pt} \settowidth{\labelwidth}{#1.}
     \sloppy
    }}{\end{list}}

\parindent=1.5pc

\renewcommand{\thefootnote}{\fnsymbol{footnote} }

\newcommand{\s}{\\ \vspace*{-2mm} }
\newcommand{\nn}{\noindent}
\newcommand{\non}{\nonumber}
\newcommand{\ee}{e^+ e^-}
\newcommand{\ra}{\rightarrow}
\newcommand{\lra}{\longrightarrow}
\newcommand{\beq}{\begin{eqnarray}}
\newcommand{\eeq}{\end{eqnarray}}
\newcommand{\tb}{{\rm tg} \beta}
\newcommand{\lsim}{\raisebox{-0.13cm}{~\shortstack{$<$ \\[-0.07cm] $\sim$}}~}
\newcommand{\gsim}{\raisebox{-0.13cm}{~\shortstack{$>$ \\[-0.07cm] $\sim$}}~}

\begin{flushright}
DESY 95--244\\
KA--TP--14--95\\
December 1995 \\
\end{flushright}

\begin{center}{{\bf HEAVY HIGGS BOSONS AT TeV e$^+$e$^-$ COLLIDERS\footnote{
Talk given by A.D. at the Workshop on Physics and Experiments with
Linear Colliders, Morioka--Appi, Japan, September 8--12 1995.}}
\vglue .4cm
{A. DJOUADI}\\
\baselineskip=14pt
{\it Institute f\"ur Theoretische Physik, Universtit\"at Karlsruhe,}\\
\baselineskip=14pt
{\it D--76128 Karlsruhe, FRG.}\\
\vglue 0.4cm
{W. KILIAN and P. OHMANN} \\
\baselineskip=14pt
{\it Deutsches Elektronen--Synchrotron DESY, D--22603 Hamburg, FRG.}\\
\vglue 0.5cm
{\tenrm ABSTRACT}}
\end{center}
{\rightskip=3pc
 \leftskip=3pc
\tenrm\baselineskip=12pt
 \noindent
We summarize the work done by the European working group on Higgs Particles
for the Workshop ``Physics with e$^+$e$^-$ Linear Colliders", Annecy--Gran
Sasso--Hamburg, Feb.--Sept. 1995. The main focus will be on the
physics possibilities at a second phase e$^+$e$^-$ linear collider with
a center of mass energy of $\sim 1.5$ TeV.
\vglue 0.6cm}

%

{\bf\noindent 1. Introduction}
\vglue 0.2cm
\baselineskip=14pt

In previous studies,$^{1,2}$ it has been shown at great details that an
$\ee$ linear collider operating in the energy range $\sqrt{s}= 300$ to
$500$ GeV with a luminosity of $\int {\cal L}=20$ fb$^{-1}$ is an ideal
machine to search for light Higgs particles.$^{3}$

In the Standard Model (SM) the whole Higgs mass range, $M_H \lsim 200$
GeV, theoretically favored by the requirement that the SM be extended up
the Grand Unification scale, $\Lambda \sim 10^{16}$ GeV, can be covered.
The search can be carried out in three different channels: the
Higgs--strahlung process $\ee \ra ZH$ and the fusion mechanisms $WW/ZZ
\ra H$. The cross sections are rather large [especially in the very
clean Higgs--strahlung process] and the properties of the Higgs boson
[in particular spin--parity quantum numbers, couplings to gauge bosons
and fermions] can be thoroughly investigated, allowing for important
tests of the Higgs mechanism.

In the Minimal Supersymmetric extension of the Standard Model (MSSM), in
which the Higgs sector is extended to comprise three neutral
$h/H$(CP=+), $A$(CP=--) and a pair of charged scalar particles $H^\pm$,
the lightest Higgs boson $h$ has a mass $M_h \lsim 140$ GeV and can be
detected in the entire MSSM parameter space either in the
Higgs--strahlung process $\ee \ra hZ$ or in the complementary mechanism
of associated production with the pseudoscalar $\ee \ra hA$. In
addition, there is a substantial area in the MSSM parameter space where
the heavier Higgs bosons can be also found; at a 500 GeV $\ee$ collider,
this is possible if the $H,A$ and $H^\pm$ masses are less than 230 GeV.
As in the case of the SM, various properties of these Higgs bosons can
be investigated.

Higher energies are required to sweep the entire mass range of the SM
Higgs particle, $M_H \lsim 1$ TeV. These high energies will also be
needed to produce and study the heavy scalar particles of extensions of
the SM, such as the MSSM, if their masses are larger than $\sim 250$
GeV. In $\ee$ collisions, a second phase with a c.m. energy up to 1.5--2
TeV would be therefore mandatory.

A study of the potential of a 1.5 TeV $\ee$ linear collider with an
integrated luminosity of $\int {\cal L}=300$ fb$^{-1}$ per annum [to
compensate for the drop in the cross sections of the interesting
processes at high energies] was undertaken in Ref.[4]. In this
contribution, we will summarize the main points of this study.

For the SM, we will concentrate on the mass range above $M_H \gsim 250$
GeV, and summarize the main production mechanisms and decay modes. The
search of New Physics through the measurement of production cross
sections, angular distributions and decay widths will be discussed. We
will focus on the case of weakly interacting New Physics at a scale
$\Lambda \gsim 1$ TeV [described in terms of an effective Lagrangian
with dimension 6 operators]; a discussion of the strongly interacting
New Physics scenario has been given in Ref.[5].

We will investigate the properties of the heavy Higgs particles of
supersymmetric extensions of the SM. We will restrict ourselves to
the minimal extension which is highly constrained since there are only
two free parameters [at tree--level]: a Higgs mass parameter [generally
$M_A$] and the ratio of the vacuum expectation values of the two doublet
fields responsible for the symmetry breaking, $\tb$ [which in Grand
Unified Supersymmetric models with b--$\tau$ Yukawa coupling unification
is forced to be either small, $\tb \sim 1.5$, or large, $\tb \sim 50$].
The various decay modes of the heavy CP--even Higgs boson $H$, the
pseudoscalar $A$ and the charged Higgs particles $H^\pm$ [in particular
the below--threshold three body decays and the decays into
supersymmetric particles] will be discussed. The main production
processes of $H,A$ and $H^\pm$ and the multiple production of the light
MSSM Higgs boson [which allows the determination of Higgs trilinear
couplings] will be summarized.

We will restrict ourselves to the $\ee$ mode of the linear collider,
other options such as $\gamma \gamma$ have been discussed elsewhere at
this Worskhop.$^6$. Some aspects related to the properties of light
Higgs particles in the SM and the MSSM, as well as the discussion of the
Higgs sector of non-minimal extensions of the SM can be found in
Ref.[4]. For more details on the topics presented in this contribution
and for a complete list of references, we refer to the report Ref.[4].

\bigskip

{\bf\noindent 2. Standard Model Higgs}
\vglue 0.2cm
\baselineskip=14pt

{\it\noindent 2.1 Production Processes and Decay Modes}
\vglue 0.2cm
\baselineskip=14pt

In the SM, the Higgs profile is uniquely determined once $M_H$ is fixed.
For a heavy Higgs boson, $M_H \gsim 250$ GeV, the decay width, the
branching ratios and the production cross sections are given by the
strength of the Yukawa couplings to top quarks and $W/Z$ gauge bosons,
the scale of which is set by the $t/W/Z$ masses.

At $\ee$ linear colliders operating in the TeV energy range, the
main production mechanisms for Higgs particles are

\begin{eqnarray}
\begin{array}{lccl}
(a)  & \ \ {\rm bremsstrahlung \ process} & \ \ \ee & \ra (Z) \ra Z+H \non \\
(b)  & \ \ WW \ {\rm fusion \ process} & \ \ \ee & \ra \bar{\nu} \ \nu \
(WW) \ra \bar{\nu} \ \nu \ + H \non \\
(c)  & \ \ ZZ \ {\rm fusion \ process} & \ \ \ee & \ra e^+ e^- (ZZ) \ra
e^+ e^- + H
\end{array}
\end{eqnarray}

\begin{figure}[htbp]
\vspace*{-.6cm}
\centerline{\psfig{figure=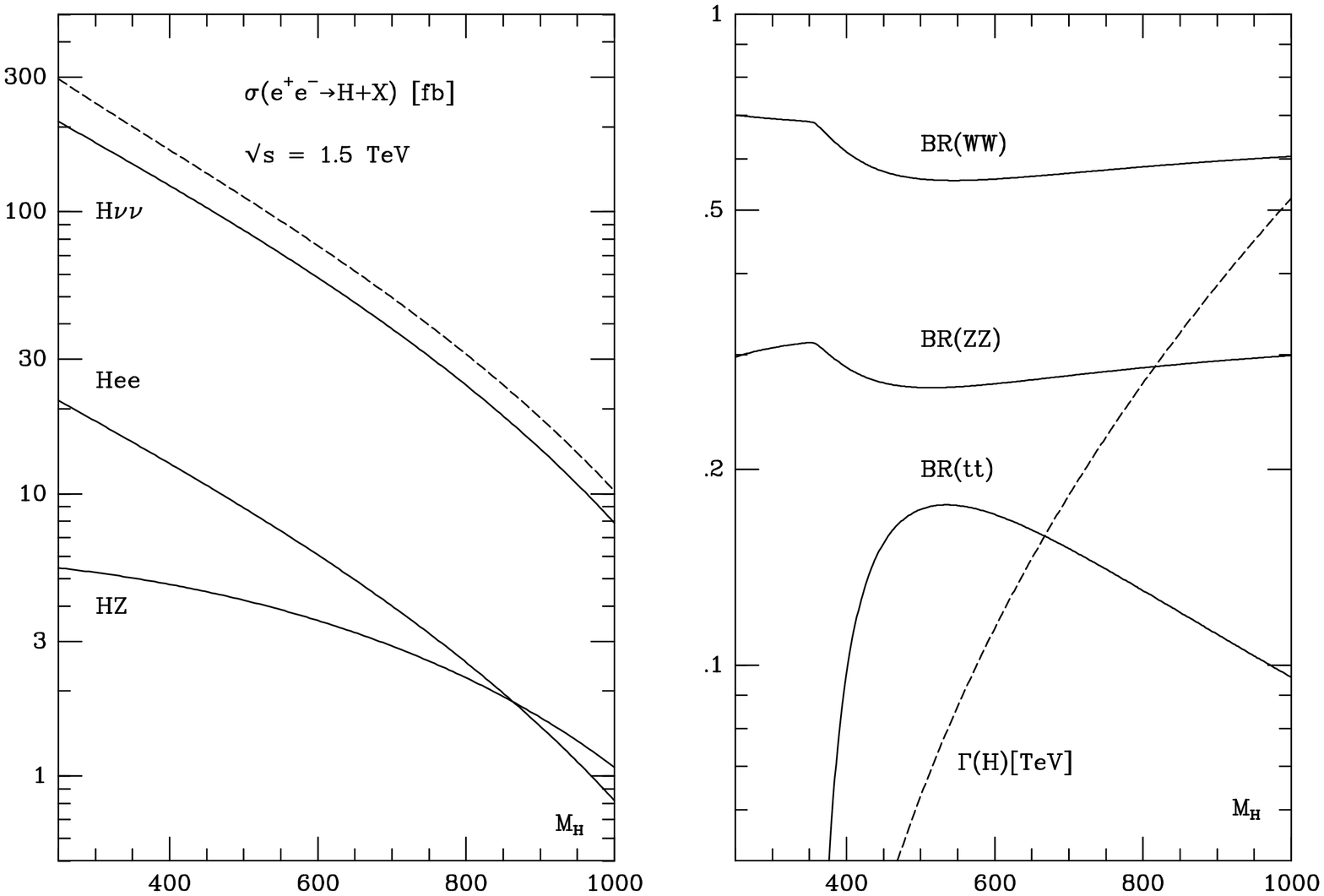,height=9.cm,angle=-90}}
\vspace*{-.6cm}
\noindent {\small Fig.1: a) Cross sections for the main production
processes of Higgs bosons at $\sqrt{s}=1.5$ TeV, and (b) the branching
ratios and the total decay width of the Higgs particle.}
\end{figure}

The cross sections are shown in Fig.1a as a function of the Higgs mass at
$\sqrt{s}=1.5$ TeV. The $WW$ fusion mechanism is by far the dominant
process since the cross section grows $\sim M_W^{-2}$log$s$. At these
energies, one can use the effective longitudinal $W$ approximation which
gives a result [dashed line] only slightly larger than the exact cross
section. The cross section is peaked in the forward and backward region
and the relevant energy scale at the $WWH$ vertex is $M_H$. Since the
NC couplings are smaller than the CC couplings, the cross section for
the $ZZ$ fusion process is $\sim 16 \cos^4 \theta_W$, i.e. one order of
magnitude smaller than for $WW$ fusion. However, the lower rate might be
partly compensated by the cleaner $H\ee$ final state.

The bremsstrahlung process which is dominant for moderate values of
$M_H/\sqrt{s}$, falls off like $\sim s^{-1}$ at high energies and the
cross section [which is peaked in the central region] is very small
compared to the dominant $WW$ fusion mechanism. However, as will be
discussed later, this process will be very useful to study the
properties of the Higgs boson since the relevant energy scale at the
$HHZ$ vertex is at $\sqrt{s}$.

For an integrated luminosity of $\int {\cal L}=300$ fb$^{-1}$, 60.000
(2.500) events in the $WW$ fusion process and 1.500 (300) in the
Higgs--strahlung process are to be expected for $M_H=250(1000)$ GeV.
Note that the longitudinal polarization of the initial beam(s) will
increase the effective $WW$ luminosity by a factor 2(4).

In the high mass range, $M_H \gsim 250$ GeV, the Higgs bosons decay
almost exclusively into $WW$ and $ZZ$ pairs, with branching ratios of
2/3 for $WW$ and 1/3 for $ZZ$; Fig.1b. The opening of the $t\bar{t}$
channel does not alter significantly this pattern, since for large Higgs
masses, the $t\bar{t}$ decay width rises only linearly with $M_H$ while
the decay widths to $W$ and $Z$ bosons grow with $M_H^3$. The latter
rise of the width into longitudinal gauge bosons makes that for heavy
Higgs bosons, the total width [dashed line] is very large and can be
measured experimentally. For $M_H \sim 1$ TeV, the total width of the
Higgs boson becomes comparable to its mass.

Once the Higgs boson has been found, it will be very important to
explore its properties.  The zero--spin of the Higgs particle is
reflected in the angular distribution in the bremsstrahl process which
asymptotically must follow the $\sin ^2\theta$ law, corresponding to the
predominantly longitudinal polarization of the accompanying $Z$ boson.
Correlations among the final fermions in the production process
$\ee \ra ZH \ra \bar{f}f \bar{t}t$ and in the decays processes $H \ra WW/ZZ
\ra$ 4 fermions, also allow to determine the CP properties of the Higgs
particle.

Of great importance is the measurement of the couplings to gauge bosons
and matter particles. The strength of the couplings to $Z$ and $W$
bosons is reflected in the magnitude of the production cross sections.
The relative strength of the couplings to top quarks is accessible
through the decay branching ratio which is at the level of 10\% for $M_H
>350$ GeV. [For smaller Higgs masses, the $ttH$ coupling can be accessed
in Higgs--strahlung off top quarks at lower energies, where the rates
are more important.]

The Higgs boson self--coupling can be measured in the processes
\begin{eqnarray}
\ee \ra Z+HH \ \ , \ \ {\rm and} \ \ WW/ZZ \ra HH
\end{eqnarray}
While the cross section is rather small in the $\ee \ra Z+HH$ process
[less than 1 fb already at $\sqrt{s}=500$ GeV for $M_H \sim 60$ GeV],
it is sizeable in the $WW$ fusion mechanism: at a 1.5 TeV collider, it
is the order of 1 fb for $M_H \lsim 200$ GeV [see next section]. With
${\cal L}=300$ fb$^{-1}$, a few hundred events can be expected and
a measurement of the trilinear coupling is possible, allowing to reconstruct
the scalar potential; a decisive test of the Higgs mechanism.

\bigskip

{\it\noindent 2.2 New Physics Effects}
\vglue 0.2cm
\baselineskip=14pt

The large sample of Higgs bosons that can be produced in the $1.5$ TeV
$e^+e^-$ mode at high luminosity allows for tests of the Standard
Model predictions at the percent level.  Thus one can search for
indirect effects of New Physics on the Higgs sector which are
associated to a scale $\Lambda$ beyond the reach of the collider for
direct particle production.  They can be parameterized in terms of
effective non-renormalizable interactions among the Higgs boson and
the other SM particles, and are suppressed by powers
of $\Lambda^{-2}$.  The largest effects are induced by dimension-six
operators when interference with the Standard Model amplitudes is
possible.  Otherwise the New Physics contributions are suppressed at
least by $\Lambda^{-4}$.

In terms of the operator basis introduced by Buchm\"uller and Wyler, the
operator $(\partial(\phi^\dagger\phi))^2$ amounts to a common
renormalization of all Higgs couplings.  A non-zero coefficient would
result in a deviation of the total Higgs production rate from the
Standard Model prediction.  Similarly, a deviation in the ratio of $ZZ$
fusion over $WW$ fusion, which in the SM is approximately equal to $0.1$
over the whole Higgs mass range, would indicate a contribution of the
operator $(D\phi)^\dagger\phi\phi^\dagger (D\phi)$.  This operator
breaks the custodial $SU_2$ symmetry of the Higgs sector; hence a
measurement of $\sigma(ZZ\to H)/\sigma(WW\to H)$ is complementary to the
measurement of the $\rho$ parameter at lower energies.  Obviously, the
branching ratios $BR(H\to WW)$ and $BR(H\to ZZ)$ are affected by the
same corrections.

Although the Higgs-strahlung process $e^+e^-\to ZH$ plays only a minor
role at high energies, it is important for the search for new
interactions, since the effective c.m.\ energy in the Higgs production
is equal to the total collider energy.  In contrast, in the fusion
processes the relevant scale is set by the Higgs mass.  The total
cross section for Higgs-strahlung is affected by contact terms $eeZH$
which are induced by operators that also modify the $eeZ$ couplings
$g_V$ and $g_A$.  The effect of the contact terms grows with
increasing c.m.\ energy, so that the limits on the effective scale
$\Lambda$ obtained by LEP can be improved by more than one order of
magnitude at a $1.5$ TeV linear collider.  On the other hand, in the
fusion processes contact terms will probably be unobservable.

An analysis of the angular distribution of Higgs bosons will give
information on anomalous couplings to transversal $W$ and $Z$ gauge
bosons.  However, according to the equivalence principle Higgs bosons
are produced mostly by fusion of longitudinal gauge bosons, and the
interference with production by transversal gauge bosons due to new
interactions is suppressed.  Similarly, anomalous couplings to photons
are difficult to observe since the process $e^+e^-\to H\gamma$ and the
decay $H\to\gamma\gamma$ occur in the Standard Model only at the loop
level. Anomalous couplings of the Higgs boson to top quarks can be
accessed through the analysis of the branching ratio which is
${\cal O}(10\%)$ when the decay is kinematically allowed.

With increasing Higgs mass ($\gsim 400$ GeV) the lower production rate
diminishes the resolving power on New Physics effects.  In particular,
the cross section for double Higgs production falls below the limit of
observability, so that the trilinear Higgs coupling is no longer
directly measurable.  On the other hand, radiative corrections to the
production and decay processes are determined by the size of the Higgs
self-coupling and become more important for a heavier Higgs boson, so
that New Physics contributions may become accessible through their
loop effects.  In order to be sensitive to these corrections it is
crucial to measure the shape of the Higgs resonance in the $WW$ and
$ZZ$ channels beyond the peak, where due to the rapidly decreasing
effective luminosity of $W$ and $Z$ bosons in the initial state the
event rate will be low.

For Higgs masses near the upper limit ($\gsim 700$ GeV) the Standard Model
predictions become increasingly unreliable, since higher-order
corrections approach the lowest-order contribution in size.  In this
mass range an effective Lagrangian with a linear Higgs
representation is no longer appropriate.  The usual approach is to use
a nonlinear representation of the symmetry-breaking sector instead,
which leaves freedom to incorporate arbitrary Higgs couplings, or to
dispense of a scalar resonance altogether.$^5$

\newpage

{\bf\noindent 3. Minimal Supersymmetric Standard Model}
\vglue 0.2cm
\baselineskip=14pt

{\it\noindent 3.1 Decay Modes}
\vglue 0.2cm
\baselineskip=14pt

The decay pattern of the SUSY Higgs bosons is determined to a large
extent by their couplings to fermions and gauge bosons; these couplings
will in general depend strongly on the angles $\alpha$ and $\beta$. The
pseudoscalar and charged Higgs boson couplings to down (up) type
fermions are (inversely) proportional to $\tb$; $A$ has no tree level
couplings to gauge bosons.  For the CP--even Higgs bosons, the couplings
to down (up) type fermions are enhanced (suppressed) compared to the SM
Higgs couplings [$\tb>1$]; the couplings to gauge bosons are suppressed
by $\sin/\cos(\beta-\alpha)$ factors.

In the following, we will summarize the various decay modes of the heavy
Higgs bosons, first in the case where the decay channels into
supersymmetric particles are shut, and then when these decays are
allowed. We will discuss the two scenarios of small and large $\tb$
values, favored by SUSY--GUT models with $b$--$\tau$ Yukawa coupling
unification. The QCD corrections to the hadronic modes have been
recently summarized [see [4] for details] and will not be discussed here.
\vglue 0.2cm
\baselineskip=14pt

\noindent \underline{Non--SUSY Two-- and Three--Body Decay modes}
\vglue 0.2cm
\baselineskip=14pt

For large values of $\tb$ the pattern is simple, a result of the strong
enhancement of the Higgs couplings  to down--type fermions. The neutral
Higgs bosons will decay into $b\bar{b}$ and $\tau^+ \tau^-$ pairs, the
charged Higgs bosons into $\tau \nu_\tau$ pairs below and $tb$ pairs
above the top--bottom threshold. For $H$, only when $M_H$ approaches
its minimal value is this simple rule modified; in this case it will
mainly decay into $hh$ and $AA$ final states.

For small values of $\tb\sim 1$ the decay pattern of the heavy neutral
Higgs bosons  is much more complicated. The $b$ decays are in general
not dominant any more; instead, cascade decays to pairs of light Higgs
bosons and mixed pairs of Higgs and gauge bosons are important.
Moreover, decays to gauge boson pairs play a major role. However, for
very large masses, the neutral Higgs bosons decay almost exclusively to
top quark pairs. The decay pattern of the charged Higgs bosons for small
$\tb$ is similar to that at large $\tb$ except in the intermediate mass
range where cascade decays to light Higgs and $W$ bosons are dominant.

\begin{figure}[htbp]
\vspace*{-.3cm}
\centerline{\psfig{figure=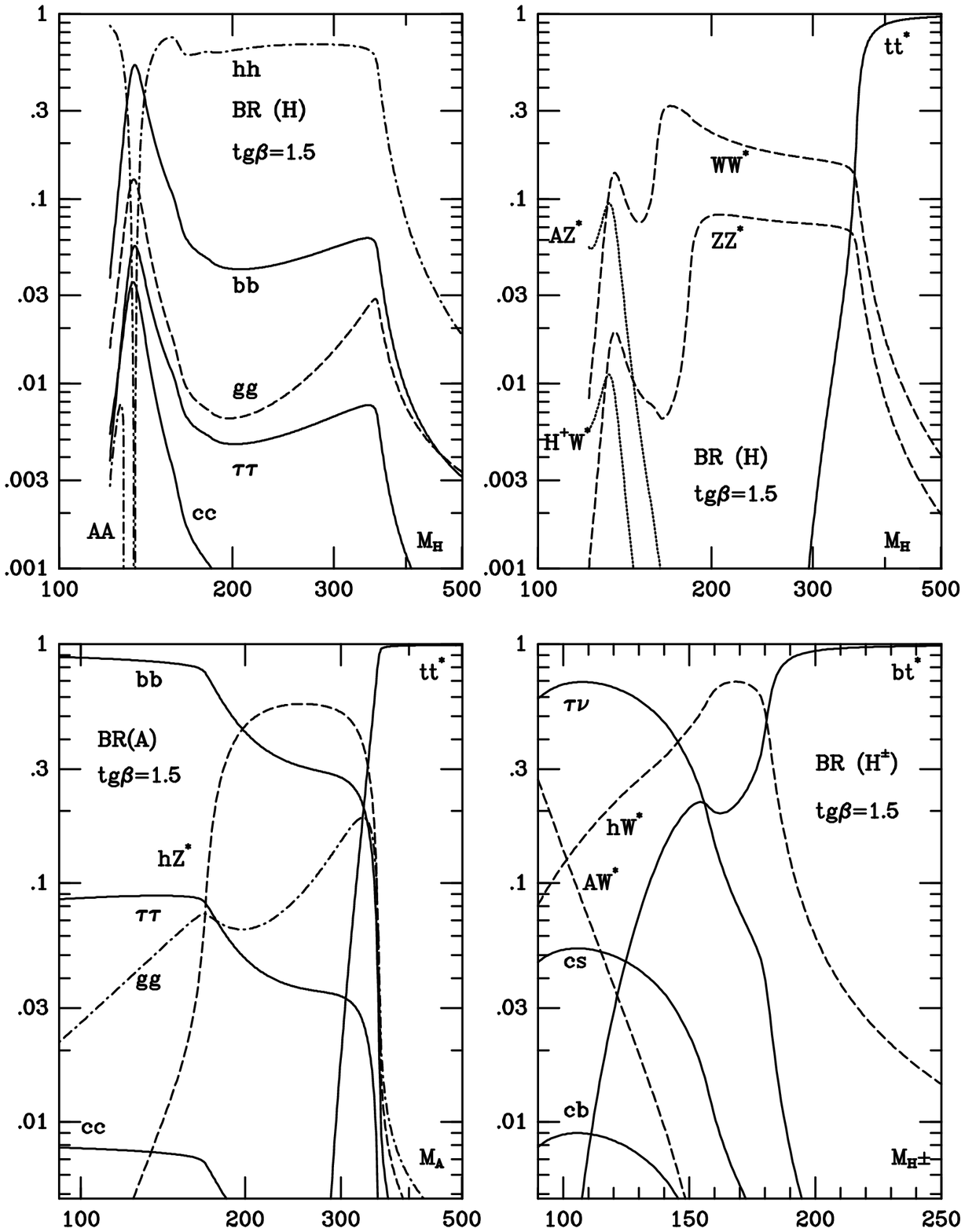,height=21.5cm}}
\vspace*{-4.cm}
\noindent {\small Fig.2: Branching ratios for the heavy CP--even, the
CP--odd and the charged Higgs bosons, including the three--body decays,
for $\tb=1.5$ and no stop mixing.}

\bigskip

The branching ratios for $h,A$ and $H^\pm$ decays are shown in Fig.2 for
$\tb=1.5$, in the case where the mixing in the stop sector is neglected.
\end{figure}

Besides these two--body decays, below--threshold three--body Higgs decay
modes can play an important role. It is well--known that SM Higgs decays
into real and virtual $Z$  pairs are quite substantial: the suppression
by the off--shell propagator and the additional $Zff$ coupling is at
least partly compensated by the large Higgs coupling to the $Z$ bosons.
For the same reason, three--body decays of MSSM Higgs particles mediated
by gauge bosons, heavy Higgs bosons and top quarks, are of physical
interest. Important three-body decays for the heavy CP--even, the
CP--odd and the charged Higgs bosons, analyzed recently are
\begin{eqnarray}
H &\ra& VV^* \ra V f \bar{f}^{(')} \ , \
        AZ^* \ra  A f \bar{f} \ , \
        H^\pm W^{\mp *} \ra  H^\pm f \bar{f}' \ , \
        \bar{t}t^* \ra \bar{t} b W^+ \\
A & \ra & hZ^* \ra h f \bar{f} \ , \
        \bar{t}t^* \ra \bar{t} bW^+  \\
H^\pm & \ra & hW^* \ra h f \bar{f}' \ , \
        AW^* \ra A f \bar{f}' \ , \
        \bar{b}t^* \ra \bar{b}bW
\end{eqnarray}

For the heavy Higgs boson $H$, the decay $H\ra hh$ is the dominant
channel, superseded by $t\bar{t}$ decays above the threshold [for the
latter, the inclusion of the three--body modes provides a smooth
transition from below to above threshold]. This rule is only broken for
Higgs masses of about 140 GeV where an accidentally small value of the
$\lambda_{Hhh}$ coupling allows the $b \bar{b}$ and $WW^*$ decay modes
to become dominant. Important channels in general, below the $t\bar{t}$
threshold, are decays to pairs of gauge bosons and $b\bar{b}$ decays.
In a restricted range of $M_H$, below--threshold $AZ^*$ and
$H^{\pm}W^{\mp *}$ also play a non--negligible role.

In the case of the pseudoscalar $A$, the dominant modes are the $A\ra
b\bar{b}$ and $A \ra t\bar{t}$ decays below the $hZ$ and $t\bar{t}$
thresholds respectively; in the intermediate mass region, $M_A=200$ to
$300$ GeV, the decay $A \ra hZ^*$ [which reaches $\sim 1\%$ already at
$M_A =130$ GeV] dominates. The gluonic decays are significant around the
$t\bar{t}$ threshold.

For the charged Higgs boson, the inclusion of the three--body decay
modes will reduce the branching ratio for the $\tau\nu$ channel quite
significantly. Indeed, this decay does not overwhelm all the other modes
since the three--body decay channels $H^+ \ra hW^*$ as well as
$H^+\ra AW^*$ in the low mass range and $H^+ \ra bt^*$ in the
intermediate mass range have appreciable branching ratios.
\vglue 0.2cm
\baselineskip=14pt

\noindent \underline{SUSY Decay modes}
\vglue 0.2cm
\baselineskip=14pt

In the previous discussion, we have assumed that decay channels into
neutralinos, charginos and sfermions are shut. However, these channels
could play a significant role, since some of these particles [at least
the lightest neutralinos, charginos, stop squarks and the sleptons] can
have masses in the ${\cal O}(100$ GeV) range or less.

To discuss these decays in the general case is an almost impossible task
because of the many parameters that one has to deal with. We have
therefore performed the analysis in the MSSM constrained by minimal
Supergravity, in which the SUSY sector is described in terms of five
universal parameters at the GUT scale: the common scalar mass $m_0$, the
common gaugino mass $M_{1/2}$, the trilinear coupling $A$, the bilinear
coupling $B$ and the higgsino mass $\mu$. These parameters evolve
according to the RGEs, forming the supersymmetric particle spectrum at
low energy.

The requirement of radiative electroweak symmetry breaking further
constrains the SUSY spectrum, since the minimization of the one--loop
Higgs potential specifies the parameter $\mu$ [to within a sign] and
also $B$. The unification of the $b$ and $\tau$ Yukawa couplings gives
another constraint: in the $\lambda_t$ fixed--point region, the value of
$\tb$ is fixed by the top quark mass through: $m_t \simeq (200~{\rm
GeV}) \sin\beta$, leading to $\tb \simeq 1.75$. There also exists a
high--$\tb$ [$\lambda_b$ and $\lambda_\tau$ fixed--point] region for
which $\tb \sim$ 50--60 [though disfavoured by $b \rightarrow s \gamma$
and Dark Matter constraints]. If one also notes that moderate values of
the trilinear coupling $A$ [e.g. $|A| \lsim 500$ GeV at the GUT scale]
have little effect on the resulting spectrum, then the whole SUSY
spectrum will be a function of only a few parameters: $\tb$ which we
take to be 1.75 and 50, the sign of $\mu$ [we will use the convention
of Haber and Kane], $m_0$ which in practice we replace with $M_A$
taking the two illustrative values $M_A =300$ and 600 GeV [note that
for these large values, $M_H \sim M_{H^\pm} \sim M_A$], and finally the
common gaugino mass $M_{1/2}$ that we will freely vary.

\begin{figure}[htbp]
\vspace*{-.3cm}
\centerline{\psfig{figure=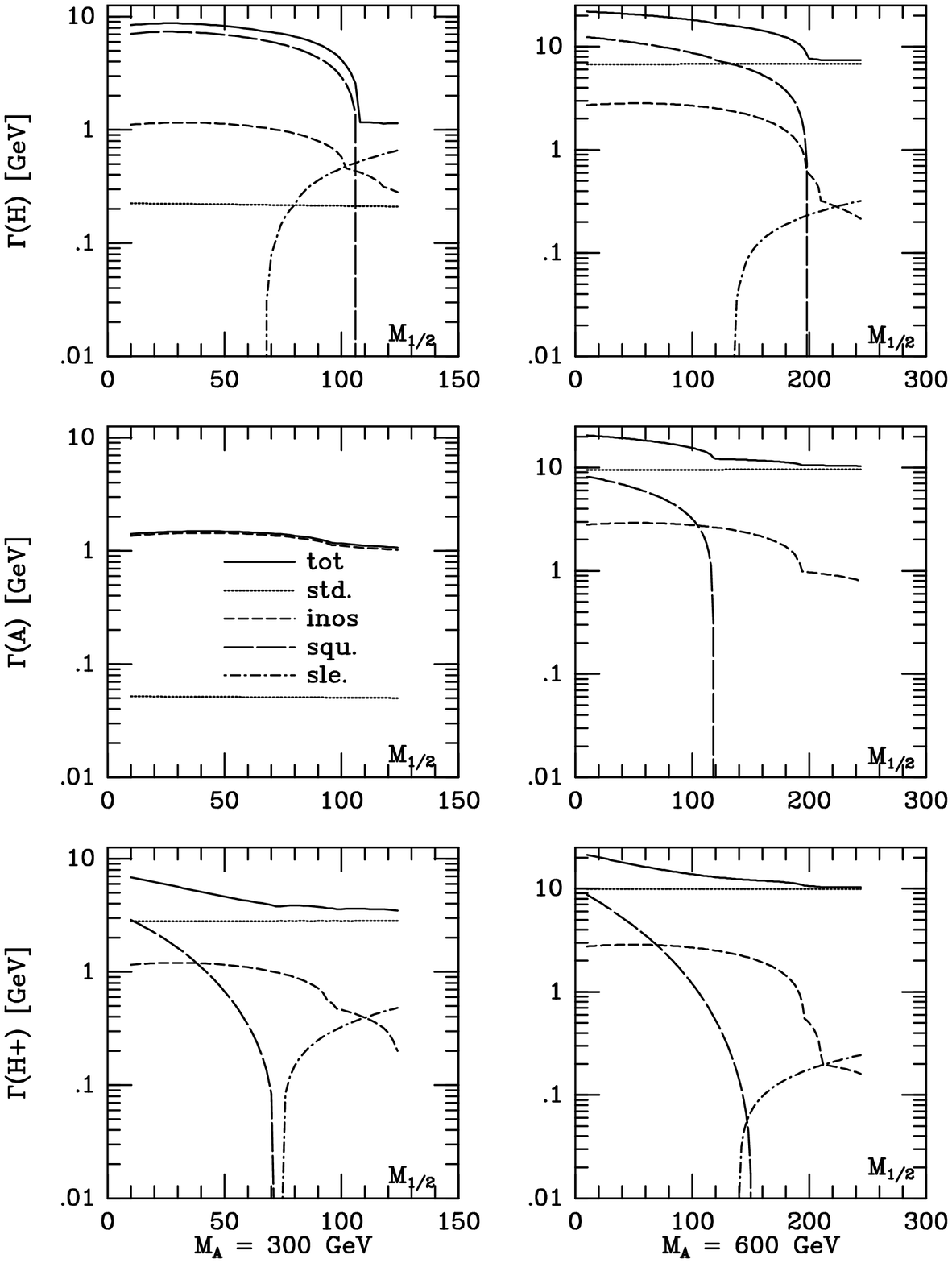,height=21cm}}
\vspace*{-2.cm}
\noindent {\small Fig.3: Decay widths for the SUSY decay modes of the
heavy CP--even, the CP--odd and the charged Higgs bosons, for $\tb=1.75$.
The total and the non--SUSY widths are also shown.}
\end{figure}

The decay widths of the heavy CP-even, the CP--odd and the charged Higgs
bosons, $H,A$ and $H^\pm$, into pairs of neutralinos and
charginos [dashed lines], squarks [long--dashed  lines] and sleptons
[dot--dashed lines], as well as the total [solid lines] and non--SUSY
[dotted--lines] decay widths, are shown in Fig.3 for $\tb=1.75$, $\mu>0$
and two values of $M_A=300$ [left curves] and $600$ GeV [right
curves].

For $M_A=300$ GeV, i.e. below the $t\bar{t}$ threshold, the widths of
the decays of $H$ into inos, sleptons and squarks are much larger than
the non--SUSY decays. In particular, squark [in fact stop and sbottom
only] decays are almost two--orders of magnitude larger when
kinematically allowed. The situation changes dramatically for larger $A$
masses when the $t\bar{t}$ channel opens up: only the decays into stop
pairs, when allowed, are competitive with the dominant $H \ra t\bar{t}$
channel. Nevertheless, the decays into inos are still substantial having
branching ratios at the level of 20\%; the decay widths into sleptons
never exceed the few percent level.

In the case of the pseudoscalar $A$, because of CP--invariance and the
fact that sfermion mixing is small except in the stop sector, only the
decays into inos and $A \ra \tilde{t}_1 \tilde{t}_2$
decays are allowed. For these channels, the situation is quite similar
to the case of $H$: below the $t\bar{t}$ threshold the decay width into
ino pairs is much larger than the non--SUSY decay widths [here
$\tilde{t}_2$ is too heavy for the $A \ra \tilde{t}_1 \tilde{t}_2$ decay
to be allowed], but above $2m_t$ only the $A \ra \tilde{t}_1
\tilde{t}_2$ channel competes with the $t\bar{t}$ decays.

For the charged Higgs boson $H^\pm$, only the decay channel $H^+ \ra
\tilde{t}_1 \tilde{b}_1$ [when kinematically allowed] competes with the
dominant decay mode $H^+\ra t\bar{b}$, yet the $\tilde{\chi}^+ \tilde{
\chi}^0$ decays have a branching ratio of a few ten percent; the
decays into sleptons are at most of the order of a few percent.

In the case where $\mu<0$, the situation is quite similar as above. For
large $\tb$ values, $\tb\sim 50$, all gauginos and sfermions are very
heavy and therefore kinematically inaccessible, except for the lightest
neutralino and the $\tau$ slepton. Moreover, the $b\bar{b}/\tau \tau$
and $t\bar{b}$/$\tau \nu$ [for the neutral and charged Higgs bosons
respectively] are enhanced so strongly, that they leave no chance for
the SUSY decay modes to be significant. Therefore, for large $\tb$, the
simple pattern of $bb/\tau\tau$ and $tb$ decays for heavy neutral and
charged Higgs bosons still holds true even when the SUSY decays are
allowed.
\vglue 0.2cm
\baselineskip=14pt

\noindent \underline{Total Decay Widths}
\vglue 0.2cm
\baselineskip=14pt

The total widths of the SUSY Higgs particles are in general considerably
smaller than the width of the SM Higgs, due to the absence or the
suppression of the decays to longitudinal $W/Z$ bosons which grow as
$G_F M_H^3$. The dominant decay modes are built-up by top quarks so that
the widths rise only linearly with the Higgs masses $\sim G_F m_t^2
M_H$. Including the SUSY modes will not change the situation
considerably since the widths still grow only linearly with the Higgs
mass. However, for large $\tb$ values, the decay widths of all the five
Higgs bosons are determined by $b$--quark final states and they scale
like ${\rm tg}^2\beta$ [except for $h$ and $H$ near their maximal and
minimal mass values, respectively]; the $H,A$ and $H^+$ widths therefore
become experimentally significant, for $\tb$ values of order $\gsim 30$
and for large Higgs masses.

\bigskip

{\it\noindent 3.2 Production Processes}
\vglue 0.2cm
\baselineskip=14pt

\noindent \underline{Main Production mechanisms}
\vglue 0.2cm
\baselineskip=14pt

At $\ee$ linear colliders operating in the TeV energy range, the main
production mechanism for the heavy neutral Higgs particles is the
associated $H$ and $A$ production
\begin{eqnarray}
\ee \ra Z^* \ra AH
\end{eqnarray}
This cross section is proportional to $\sin^2(\beta-\alpha)$ which is
close to one near the decoupling limit. If $H$ and $A$ are very heavy,
the cross sections for the Higgs--strahlung process $\ee \ra HZ$, the
fusion processes $\ee \ra W^*W^* (Z^* Z^*) \ra H \nu_e \bar{\nu}_e
(\ee)$ and the pair production process $\ee \ra hA$ are very small since
they are suppressed by a factor $\cos^2 (\alpha-\beta)$ which is then
close to zero. These cross sections are significant only for $H$
masses below 400 GeV and small values of $\tb \sim 1.5$.

\begin{figure}[htbp]
\vspace*{-.6cm}
\centerline{\psfig{figure=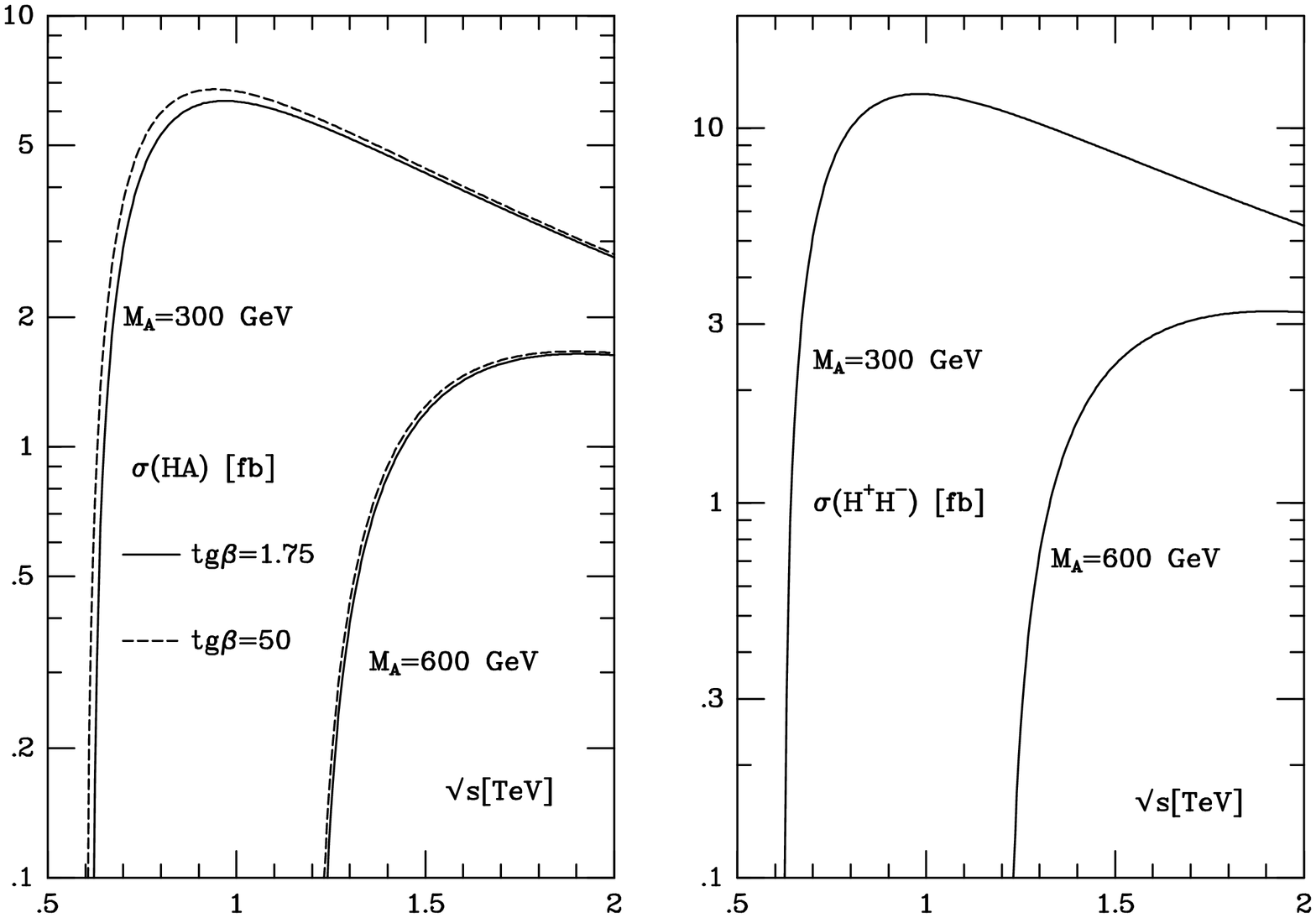,height=9.cm,angle=-90}}
\vspace*{-.6cm}
\noindent {\small Fig.4: Cross sections for $\ee \ra HA$ and $\ee \ra H^+H^-$
as a function of the c.m. energy.}
\end{figure}

For charged Higgs particles, the production process [the cross section
of which does not depend on any parameter other than $M_{H^\pm}$]
proceeds through virtual photon and $Z$--boson exchange
\begin{eqnarray}
\ee \ra \gamma^* , Z^* \ra H^+ H^-
\end{eqnarray}

The cross sections for the $(AH)$ and $(H^+ H^-)$ processes are shown in
Fig.4 as a function of the c.m. energy for $M_A [\simeq M_H \simeq
M_{H^\pm}] =300$ and 600 GeV. [In the case of $(AH)$ two values of $\tb$
are shown; but as can be seen, the cross section depend only slightly on
the value of $\tb$, since we are close to the decoupling limit]. Since
the processes are mediated by $s$--channel exchanges, the cross sections
scale like $1/s$ sufficiently above the thresholds. Due to the presence
of the additional photon channel, the cross section for $(H^+H^-)$ is
twice as large as for $(AH)$. At $\sqrt{s} \sim 1.5$ TeV, the cross
sections are of the order of a few fb: for a luminosity ${\cal L}=300$
fb$^{-1}$/year, 500 to 1000 events could be expected. For masses above
350 GeV and in the absence of SUSY decays, the final states consist of
$b\bar{b}b\bar{b}$ [for large $\tb$] and $t\bar{t}t\bar{t}$ [for small
$\tb$] in the case of $(HA)$ production and $b\bar{b}t\bar{t}$ in the
case of $(H^+ H^-)$ production. Efficient $b$--tagging through
$\mu$-vertexing is therefore required to separate the signals from the
backgrounds. Rich final states maybe be investigated if the decay
channels into SUSY particle are open.
\vglue 0.2cm
\baselineskip=14pt

\noindent \underline{Multiple Higgs Production}
\vglue 0.2cm
\baselineskip=14pt

One of the main motivations to proceed to higher c.m. energies is the
availability of enough phase space to produce Higgs particles in
association with heavy states. For instance, at sufficiently high
energies, one can have multiple Higgs production, the cross sections of
which allow the trilinear couplings among the Higgs particles to be
determined. By measuring these trilinear couplings the scalar potential
can be reconstructed, allowing a fundamental test of the Higgs
mechanism. The multiple production of the lightest SUSY Higgs boson has
been analyzed recently [see [4] for details]; the results are summarized
below.

The most copious source of multiple light Higgs boson final states is
the cascade decay $H \ra hh$ where the heavy CP--even neutral Higgs
boson is produced either by Higgs--strahlung and associated pair
production, or in the $WW(ZZ)$ fusion mechanisms. The cross sections are
sizeable for small $\tb$ values and $H$ masses below $\sim 400$ GeV, and
in this range the $H\ra hh$ branching ratio is neither too small nor too
close to unity [Fig.2] if the SUSY decay channels are closed. Apart
from small mass intervals, the other important decay modes are
$WW^*/ZZ^*$ decays; since the $HVV$ couplings can be measured through
the production cross sections of the fusion and Higgs--strahlung
processes, the branching ratio BR$(H \ra hh)$ can be exploited to
measure the coupling $\lambda_{Hhh}$.

Besides the previous process, multiple light Higgs bosons $h$ can [in
principle] be generated in the MSSM by three mechanisms: (i) double
Higgs--strahlung in the continuum, with a final state $Z$ boson $\ee \ra
hhZ$; (ii) associated production with the pseudoscalar $A$ in the
continuum, $\ee \ra hhA$; and (iii) non--resonant $WW$ fusion in the
continuum, $\ee \ra \bar{\nu}_e \nu_e hh$. These processes are
disfavoured by an additional power of the electroweak coupling compared
to the resonance processes; nevertheless, they must be analyzed in order
to measure the values of the $hhh$ and $hAA$ couplings. The situation
for double Higgs--strahlung and $WW$ fusion is analogous to the SM,
while the process $Ahh$ is novel.

The cross section $\sigma(\ee \ra hhZ)$ is shown for $\sqrt{s}=500$ GeV
at $\tb=1.5$ as a function of $M_h$ in Fig.5a. For small masses, the
cross section is built--up almost exclusively by $H \ra hh$ decays
[dashed curve], except close to the point where $\lambda_{Hhh} \sim 0$.
For intermediate masses, the resonance contribution is reduced and, in
particular above 90 GeV where the decoupling limit will be approached,
the continuum $hh$ production becomes dominant, finally falling down to
the SM cross section [dotted line]. After subtracting $hh$ decays, the
continuum cross section is of the same order, $\sim 0.5$ fb as in the
SM. Very high luminosity is therefore needed to measure the trilinear
$hhh$ coupling. At higher energies, since the cross section scales like
$1/s$, the rates are smaller. Prospects are similar for large $\tb$
values. The cascade decay $H \ra hh$ is restricted to a range $M_h \lsim
70$ GeV, with a cross section of $\sim 20$ fb at $\sqrt{s}=500$ GeV and
$\sim 3$ fb at 1.5 TeV. The continuum cross sections are of the order of
$0.1$ fb at both energies, so that very high luminosities will be
needed.

\begin{figure}[htbp]
\vspace*{-.6cm}
\centerline{\psfig{figure=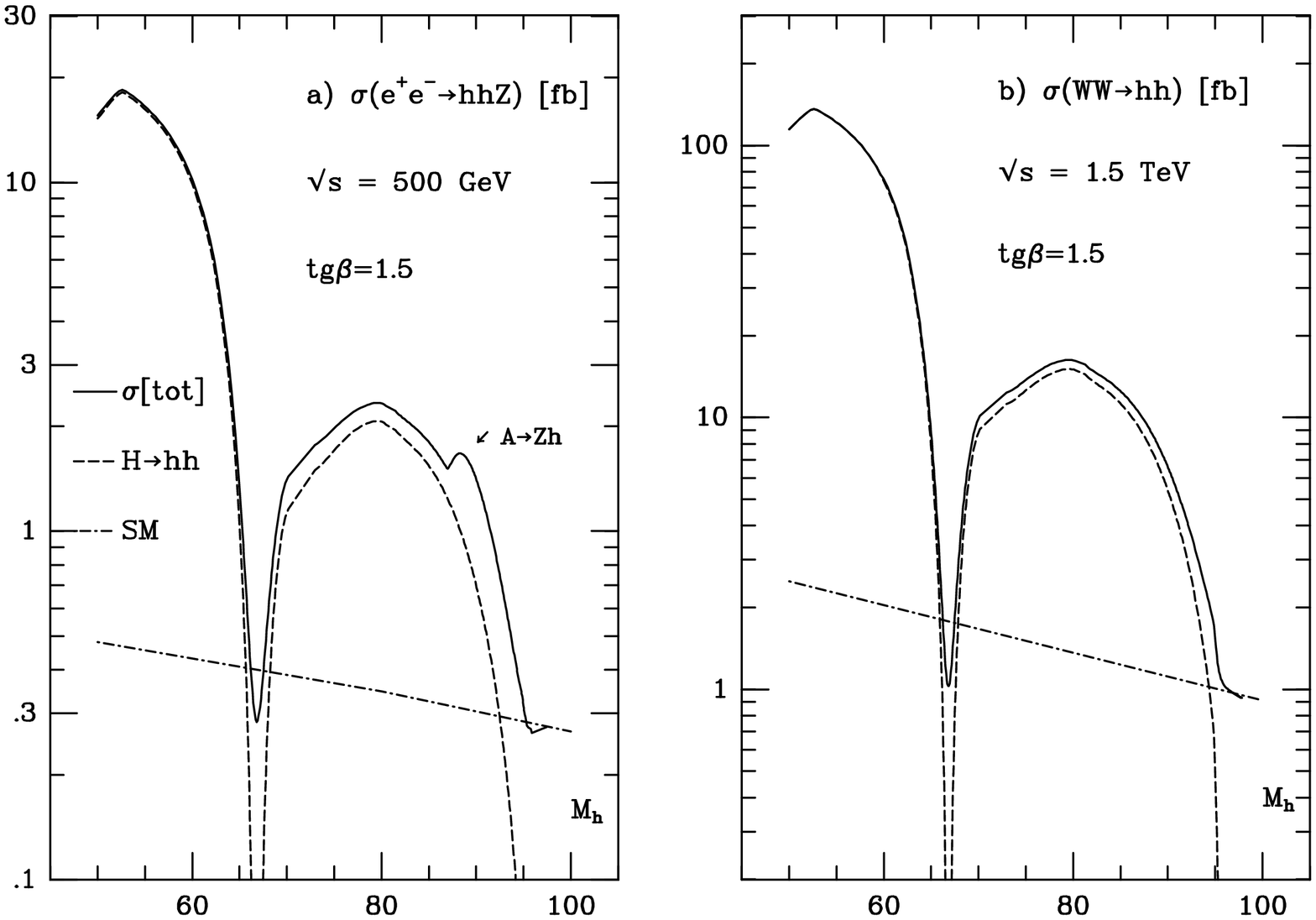,height=9cm,angle=-90}}
\vspace*{-.6cm}
\noindent {\small Fig.5: Cross sections for double Higgs production as a
function of $M_h$ at $\tb=1.5$: a) $\ee \ra hhZ$ at $\sqrt{s}=500$ GeV,
and (b) $WW \ra hh$ at $\sqrt{s}=1.5$ TeV.}
\end{figure}

The analysis has been repeated for the continuum process $\ee \ra Ahh$.
However, it turned out that the cross section is built--up almost
exclusively by the resonant process $\ee \ra AH \ra Ahh$, with a very
small continuum contribution, so that the coupling $\lambda_{hAA}$
cannot be measured in this process.

The total cross section $\sigma(\ee \ra \nu_e \bar{\nu}_e hh)$ [obtained
in the effective longitudinal $W$ approximation] is shown in Fig.~5b as
a function of $M_h$ for $\tb=1.5$ at $\sqrt{s} =1.5$ TeV. It is
significantly larger than for double Higgs--strahlung in the continuum.
Again, for very light Higgs masses, most of the events are $H \ra hh$
decays [dashed line]. The continuum $hh$ production is of the same size
as pair production of SM Higgs bosons [dotted line] which as
anticipated, is being approached near the upper limit of the $h$ mass in
the decoupling limit. The size of the continuum $hh$ fusion cross
section renders this channel more promising than double Higgs--strahlung
for the measurement of the trilinear $hhh$ coupling. For large $\tb$
values, strong destructive interference effects reduce the cross section
in the continuum to very small values, of order 10$^{-2}$ fb, before the
SM cross section is reached again in the decoupling limit. As before,
the $hh$ final state is almost exclusively built--up by the
$H\ra hh$ decays.

\bigskip

{\bf\noindent 4. Summary}
\vglue 0.2cm
\baselineskip=14pt

$\ee$ linear colliders with center of mass energies larger than $\sim
500$ GeV will be needed to search for Higgs particles in the mass range
above $M_H \sim 250$ GeV. We have summarized the rather rich potential
of an $\ee$ linear collider operating at a center of mass energy of
$\sqrt{s} \sim 1.5$ TeV.

The search for the SM Higgs particle can be extended to the entire mass
range below ${\cal O}$(1 TeV). This search can be carried out in several
complementary channels. The clean environment of the collider allows to
investigate thoroughly the properties of the Higgs boson: the mass and
decay widths can be precisely measured and the measurement of its
couplings to gauge bosons and heavy fermions as well as its
self--coupling can be performed. These measurements will allow a
stringent test of the Higgs mechanism, and provide a window for new
phenomena beyond the SM.

In Supersymmetric extensions of the SM, one would have access to the
heavy Higgs particles. In the MSSM, the heavy CP even, the pseudoscalar
and the charged Higgs bosons with masses up to the beam energy can be
found if the luminosity is high--enough, $\int {\cal L} \gsim 100$
fb$^{-1}$ at $\sqrt{s}=1.5$ TeV. The measurement of production cross
sections [including the cross sections for multiple Higgs production],
branching ratios [especially in the case where decay modes into
supersymmetric particles are present] allows fundamental tests of this
extension of the SM.

\bigskip
{\bf\noindent Acknowledgements:}
\vglue 0.1cm
\baselineskip=14pt

A.D. would like to thank the organizers of the Workshop for their
support and for creating a very stimulating atmosphere. Discussions with
H.E. Haber, R. van Kooten, Y. Okada and P. M. Zerwas are gratefully
acknowledged. A.D. is supported by Deutsche Forschungsgemeinschaft (DFG)
and P.O. by the Von Humboldt foundation.

\vglue 0.5cm
{\bf\noindent References \hfil}
\vglue 0.3cm


\begin{thebibliography}{9}

\bibitem{R1} For summaries, see H.E. Haber, Proc. ``Physics and
Experiments with Linear Colliders", Saariselk\"a 1991; F. Zwirner, {\it
ibid};  S. Komamiya, {\it ibid}; J.F. Gunion, Proc. ``Physics and
Experiments with Linear Colliders", Waikoloa, 1993; P. Janot, {\it
ibid}; Y. Okada, these proceedings; R. van Kooten, {\it ibid}.
\bibitem{R2} Proceedings of the Workshop ``$\ee$ Collisions at 500 GeV: the
Physics Potential", DESY Report 92--123, P.~Zerwas ed.
\bibitem{R3} For reviews on Higgs Physics, see J. Gunion, H. Haber, G.
Kane and S. Dawson, The Higgs Hunter's Guide, Addison--Wesley, Reading
1990; P.M. Zerwas, Proceedings of the International Conference on
High--Energy Physics, Marseille, 1993; A. Djouadi, Int. J. Mod. Phys.
A10 (1995) 1.
\bibitem{R4} A. Djouadi, H.E. Haber, P. Igo--Kemenes, P. Janot and
P.M. Zerwas [conv.] et al., Proceedings of the European Workshop
``Physics with $\ee$ Linear Colliders", Annecy--Gran Sasso--Hamburg, 1995.
\bibitem{R5} Talk given by T. Barklow, these proceedings.
\bibitem{R6} Talk given by D. Miller, these proceedings.

\end{thebibliography}
\end{document}